# ARTIFICIAL INTELLIGENCE AND ITS IMPACT ON THE FOURTH INDUSTRIAL REVOLUTION: A REVIEW

Gissel Velarde

**ABSTRACT**

*Artificial Intelligence may revolutionize everything during the so-called fourth industrial revolution, which carries several emerging technologies and could progress without precedents in human history due to its speed and scope. Government, academia, industry, and civil society show interest in understanding the multidimensional impact of the emerging industrial revolution; however, its development is hard to predict. Experts consider emerging technologies could bring tremendous benefits to humanity; at the same time, they could pose an existential risk. This paper reviews the development and trends in AI, as well as the benefits, risks, and strategies in the field. During the course of the emerging industrial revolution, the common good may be achieved in a collaborative environment of shared interests and the hardest work will be the implementation and monitoring of projects at a global scale.*

## 1. INTRODUCTION

Artificial Intelligence (AI) may revolutionize everything during the forthcoming industrial revolution, serving a variety of emerging technologies [1, 2]. Recently, after a turbulent history of successes and failures, intelligent machines demonstrate significant advances in tasks involving perception, creativity and complex strategic execution [3, 4, 5, 6, 7].

Some argue that the widespread introduction of AI technologies may cause massive job reductions and greater wealth inequality; however, unemployment decreased, and productivity increased during the previous industrial and digital revolutions [2]. Others discuss whether machines could excel human intelligence and, if so, what the consequences would be. The views on how AI will impact our future are diverse, but the main trends are four: optimistic, pessimistic, pragmatic, and doubtful [2]. The optimists predict a utopian future of unlimited wealth, with computers and human-brains connected to the cloud in a symbiotic enhancement, where humans work only on tasks of their interest and robots take care of the rest of the jobs [2]. On the contrary, the pessimists believe that optimists underestimate the danger of superintelligence leading and making important decisions for us [2]. Superintelligent machines, those smarter than the brightest people, could recursively improve themselves and outperform humans on cognitive tasks, becoming potentially dangerous if their goals attempt life [8]. The pragmatists believe that AI can be controlled through effective regulations and research on human intelligence augmentation [2]. In this category, the view of economists is that AI will make predictions cheaper, faster, better, and the complements to prediction such as decisionmaking, data-collection, or judgment will increase their value and remain as exclusive tasks to humans [9, Ch. 7]. At the same time, practitioners are aware that safety, ethics, and the alignment of goals between superintelligent machines and humanity have to be addressed soon [10]. Finally, the doubters do not believe AI can surpass biological intelligence, nor should AI be seen as a threat for humans [2].

The forthcoming industrial revolution could progress without precedents in human history due to its speed and scope, with AI playing a ubiquitous role [1, 2]. Section 2 presents the development and trends in AI. Section 3, discusses the impact of AI, its benefits, risks, strategies and the initiatives that might help drive the progress of the emerging industrial revolution towards the common good. Section 4 presents the conclusion.

## 2. THE DEVELOPMENT OF AI, MACHINE LEARNING AND DEEP LEARNING

AI is the study of enabling entities to perceive, reason and act properly and with foresight in their environment [6, 11, p. 13, p. 5]. In the middle of the nineteenth century, Ada Lovelace realized that analytical machines had the power to solve problems of complex nature [12]. Almost one century later, during the emerging time of computers, Alan Turing proposed that universal machines could do anything, even mimicking human intellect [6, p. 61]. Computers mark the beginning of the digital revolution [2].

Machine learning, a subfield in AI, is a data-driven approach that finds approximate solutions by process of generalization. In 1959, the game of checkers was used to demonstrate that machine learning enables computers to learn from experience eliminating the effort of programming them in detail [13]. Soon after, machines demonstrated to be better than humans at solving intellectually difficult tasks, such as solving problems described by lists of mathematical rules, but remained challenged for decades on intuitive tasks performed easily by humans, such as recognizing faces or speech [3, 4].

In 1989, a convolution neural network demonstrated to robustly classify handwritten digits, modeling neural activation and processing [3]. Convolutional networks symbolize the beginning of the AI revolution [2]. In 2012, deep learning machines broke by far records on the ImageNet classification benchmark [7]. In the following years, deep networks won all competitions of the ImageNet and became popular in various fields of research [3, p. 24, 7, p. 98].

Deep learning is a subfield in machine learning. The difference between shallow and deep learners is the depth of possibly learnable causal links between actions and effects [7, p. 85]. Deep models learn a hierarchy of concepts from simple to complex ones in a layered fashion and have been successful thanks to developments in learning algorithms, implementations on Graphical Processing Units (GPUs) and large amounts of data [3, 7]. Although deep learning appears as a new research frontier [4], its roots can be traced back to the 18th century, with convolution principles formally described by Lacroix in 1754 [14] and variants of linear regression methods developed in the early 1800s by Gauss and Legendre [7, p. 90].

### 2.1. Trends in AI

AI is considered as the "new electricity" of the emerging industrial revolution, and particularly deep learning will transform most industries making it indispensable to a wide range of businesses and organizations [15]. Narrow AI, the type of AI that addresses specific problems, advances rapidly in applications such as self-driving cars, games, and robotics. In contrast, Artificial General Intelligence (AGI), that intelligence outperforming humans across a range of cognitive tasks, is seen as science fiction; still, some experts consider AGI could occur by 2075 [16].

AI research and its applications expand rapidly. Some trends are identified as follows.

**2.1.1. Supervised Learning and Recommendation Systems**

Possibly the most profitable applications in AI are those making use of supervised learning algorithms: (i) predicting whether a user will click on an online advertisement, (ii) automatic object recognition in applications to image, speech, audio, text, and (iii) suggesting whether someone should be granted a loan [15]. Predicting personality and behavior from the user's digital footprints (e.g., traceable actions, communications, profile pictures) is essential for recommendation systems [17]. Recommendation systems are considered as an emerging application, possibly due to their usefulness in retail stores to suggest items that users might be interested in [18].

**2.1.2. Deep Reinforcement Learning**

Reinforcement learning agents should learn how to interact with unknown and dynamic environments, aiming at maximizing a cumulative reward function that chooses between better or worse actions [7]. Reinforcement learning has been a recurrent research topic since the 1950s [7]. In recent years, it gained extreme attention thanks to a deep reinforcement learning-based agent called AlphaGo, able to learn how to play the strategy board game Go from scratch without human knowledge [5]. AlphaGo defeated 60 professional players and a world champion after 21 days of training with an algorithm considered to generalize to find solutions on any domain [19].

**2.1.3. Representation learning**

Computational methods strongly depend on data representations. The success of machine learning algorithms seems strongly related to how discriminative are some representations in the variability of the data [20]. Representation learning has become a relevant research field within AI [20, 3, p. 1798, p. 9]. The trend is to avoid feature engineering and use end-to-end deep learning to extract relevant features directly from the rough data. However, in some applications, end-to-end approaches do not outperform some standard techniques, i.e., spectrograms are still widely used as an initial layer in audio music applications [21].

**2.1.4. Computational Creativity**

Intelligence, perception and creativity are related [22, 23], and creativity is harder to define and evaluate empirically than intelligence [23, p. 184]. Two decades ago, Boden considered creativity as a fundamental human feature and a challenge for machines; and predicted that intelligent machines would have less difficulty in becoming creative than in automating their evaluation [24]. Boden's prediction is certain so far and logical, given that in general, evaluating creativity is harder than evaluating intelligence. In computational creativity research, evaluation is non-trivial and a 'Grand Challenge' [25], despite several computational creativity evaluation methodologies and frameworks [24, 26].

Applications of deep generative models are varied. For example, generative models have been used to write automatically fluent and coherent Wikipedia articles with relevant factual information [27]; machine-created images are auctioned in art galleries for almost $100,000 [28]; and people are not better than random guessing when discriminating between pieces of music composed by J.S. Bach or by a machine [29].

**2.1.5. Human-Computer interaction**

Computational creativity can boost human intelligence by providing tools that help to reason, e.g., machines that can learn to describe complex phenomena automatically [30]. Creative machines can also be used to complete tasks such as automatically generating images from human sketches [30].

Besides, it is envisioned that robots will increasingly interact in societies, and will have to behave appropriately according to social values, norms, and legal rules [31].

### 2.1.6. Ethics, bias, fairness, privacy and safety

When designing intelligent machines, it is essential to consider addressing ethics, bias, fairness, privacy, and safety. Algorithms can potentially be biased towards features present or inferred from the data and contribute to discrimination in civil rights, financial services, or employment [32, p. 2]. Addressing privacy and fairness involves designing laws, regulations, and working towards algorithmic transparency and algorithm output auditing [32, p. 12 and 25].

Poorly designed intelligent systems present the risk of accidents in real-world applications [10]. Safe AI systems should (1) avoid adverse side effects and reward function Hacking; and (2) ensure scalable supervision, safe exploration, and robustness to work properly on environments for which the machines were not designed [10].

International organizations are working to develop and promote AI ethical guidelines. The highlevel expert group on AI of the European Commission published ethics guidelines for trustworthy AI considering regulations, laws, ethics, and robustness of systems from a technical and social perspective [33]. Similarly, The Institute of Electrical and Electronics Engineers, Incorporated (IEEE) provided ethically aligned design for autonomous and intelligent systems with recommendations on standardization, certification, and regulation towards alignment of systems and societal well-being [34]. Likewise, the Association for Computer Machinery published their Code of ethics, considering: social impact, fairness, trust, privacy, and auditability, among other principles relevant for computing professionals [35].

## 2.2. Machine Consciousness

There are various theories about consciousness and whether machines will ever be conscious is a controversial topic [36]. Consciousness refers to a degree of "awakeness" and the idea that it is a brain process became mainstream in 1994, when neuroscientists observed that if intelligence or creativity could be simulated computationally, then consciousness could too [23, p. 183]. Still, consciousness remains a mystery in neuroscience [23, p. 183]. According to a purely computational theory, machines could be regarded as conscious if they would be able to develop (1) "mental" representations globally available to the organism and (2) self-monitoring [36].

Conscious machines could prefer means or goals in conflict with those preserving humanity. Machine consciousness research and development will have to involve machine compassion inherently. Compassion is a helping behavior based on empathy connected to an imagined feeling of others [37, p. 24-25].

## 3. THE IMPACT OF AI

The fourth industrial revolution is considered as an event without precedents in human history due to its speed and scope [1]. Makridakis predicts that the forthcoming AI-powered revolution will come into full force within the next twenty years, probably impacting society and firms at a larger scale than the previous industrial ~(1712-1919) and digital revolutions ~(1943-1993) together [2]. If the future world will be utopian or dystopian is uncertain [1, 2].

There is a boom in the number of scientific discoveries, areas of application and number of emerging technologies such as biotechnology, 3-D printing, blockchain, virtual and augmented reality, internet of things, smart cities, driverless cars, robotics and AI [1]. AI is expected to affect all industries and companies, enabling extensive organizational interaction and global competition [2]. Schwab proposes that it is our responsibility to establish common values and policies that will enable opportunities for all [1, p. 17].

### 3.1. AI Benefits, Risks and Strategies

Experts consider that AI-based systems can bring us closer to solve even the most challenging problems, such as eradicating war and poverty [38, 19]. At the same time, the misuse of powerful technologies could result in catastrophic consequences. Since 2015, over 4,000 AI and robotics researchers and more than 26,000 other endorsers have signed an open letter to prevent a superiority competition of autonomous weapons [39].

Six international organizations and almost 30 countries have established AI strategies and policies since 2015 [40]. The international collaborations include the United Nations, the European Commission, the Nordic-Baltic Region, the agreement between the United Arab Emirates and India, the International Study Group of AI, and the leaders of the G7. The collaborative strategies address competitiveness and collaboration, ethics, social benefit, safety, legal regulation, innovation, and economic growth, among other topics, while the single-nation strategies are diverse.

### 3.2. The Common Good

Political authorities seem incapable of achieving the common good [41, p. 13], which refers to shared interests, values, and facilities among members of a community [42]. Various societies around the globe face several challenges reflecting deficient economic and social systems [43, 41, 44]. Schwab suggests bringing together leaders from industry, government, faith, academy, and civil society to collaborate, develop and implement sustainable solutions [1, p. 99]. In contrast, an award-winning proposal submitted to the Global Challenges Foundation proposes to decentralize global governance with blockchain and AI technologies through bottom-up participation and deliberative models [45]. Both ideas aim to promote collaboration and shared responsibility to achieve the common good globally.

Since 2017, the AI for Good Global Summit gathers government, industry, academia, and civil society; and promotes AI-based projects targeting and monitoring the Sustainable Development Goals established by the United Nations. Another effort is the Beneficial Artificial General Intelligence Conference held in 2015, 2017 and 2019. Incrementally, AI and machine learning conferences are dedicating a session on beneficial AI. Established organizations include Google's AI for Social Good, Open AI, Partnership on AI, and AI for Humanity from the Université de Montréal.

During the AI for Good Global Summit 2019, experts recognized an opportunity to achieve the Sustainable Development Goals established by the United Nations, if AI-based projects are well defined and focused on solving real problems. On the other hand, it could be possible that the gap between AI-empowered economies and non-AI-empowered economies will widen. There are several challenges such as job transformation, re-skilling workers, and creating an inclusive environment. Moreover, large repositories of data are concentrated in a few hands. Other topics discussed include the necessity to ensure trust, privacy, security, fairness, and explainability, as well as enforcing ethical guidelines, e.g., by law. However, it seems that technological innovations occur at a faster pace than policymaking.

Often decision-makers remain caught into traditional thinking or are busy with immediate issues [1], or may not fully understand the implications of a particular topic. The cyclical process of policymaking can last several years; its impact is hard to predict and has a complex and interactive environment between change-demandants, decision-makers, and the ones affected by policies [46].

Decision support systems and scientific results are often not used in policy making [47]. Research designed for social robots on decision-making, task planning, conflict resolution, and normative-goal-directed reasoning under uncertainty [31], can give light on how to use technology to support a shared social responsibility at local and global levels. The hardest work to be addressed will be implementing and monitoring collaborative projects globally [2].

## 4. CONCLUSION

AI will be a ubiquitous technology during the forthcoming industrial revolution, since it enables entities and processes to become smart. Organizations and economies adopting AI strategically, will enjoy a competitive advantage over those who do not incorporate this technology timely and adequately. Education and soft-skills development will play an essential chapter in AI strategies. In the coming years, deep learning will remain popular in AI research. AI will be applied incrementally in every research field and industry, producing substantial improvements. Still, the views on how AI will impact society and firms will remain controversial, similarly to the opinions on whether AI will outperform biological intelligence. The fourth industrial revolution promises great benefits, but entails massive challenges and risks. It seems plausible but remote to achieve the common good globally, as this will require global collaboration and shared interests.

International Journal of Artificial Intelligence & Applications (IJAIA) Vol.10, No.6, November 2019

**Authors**

**Gissel Velarde** holds a PhD degree in Computer Science and Engineering from Aalborg University. She participated as a research member of the European project, "Learning to Create" (Lrn2Cre8) of the European Commission. She has developed machine learning models for classification, structural analysis, pattern discovery, representation learning and recommendation systems.

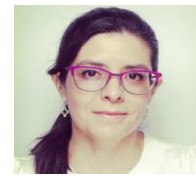